\documentclass[journal=jcisd8,manuscript=article]{achemso}

\usepackage[ruled]{algorithm2e}

\usepackage[version=3]{mhchem} 

\usepackage[utf8]{inputenc} 
\usepackage[T1]{fontenc}    
\usepackage{hyperref}       
\usepackage{url}            
\usepackage{booktabs}       
\usepackage{amsmath, amsfonts}       
\usepackage{nicefrac}       
\usepackage{microtype}      
\usepackage{lipsum}		
\usepackage{graphicx}

\usepackage{doi}
\usepackage{xcolor}

\title{Ice Phase Classification Made Easy with Score-based Denoising}

\author{Hong Sun}
\author{Sebastien Hamel}
\author{Tim Hsu}
\author{Babak Sadigh}
\author{Vince Lordi}
\email{lordi2@llnl.gov}
\author{Fei Zhou}
\email{zhou6@llnl.gov}
\affiliation{Lawrence Livermore National Laboratory, Livermore, CA 94550}

\begin{document}

	\begin{abstract}
		
		Accurate identification of ice phases is essential for understanding various physicochemical phenomena. However, such classification for structures simulated with molecular dynamics is complicated by the complex symmetries of ice polymorphs and thermal fluctuations. For this purpose, both traditional order parameters and data-driven machine learning approaches have been employed, but they often rely on expert intuition, specific geometric information, or large training datasets. In this work, we present an unsupervised phase classification framework that combines a score-based denoiser model with a subsequent model-free classification method to accurately identify ice phases. The denoiser model is trained on perturbed synthetic data of ideal reference structures, eliminating the need for large datasets and labeling efforts. The classification step utilizes the Smooth Overlap of Atomic Positions (SOAP) descriptors as the atomic fingerprint, ensuring Euclidean symmetries and transferability to various structural systems. Our approach achieves a remarkable 100\

	\end{abstract}

	\section{Introduction}
	
	Understanding water's phase behavior is crucial for investigating a wide range of physicochemical phenomena, including ice precipitation in clouds \cite{moberg2019}, CO$_{2}$ reduction \cite{asadi2016nanostructured}, water desalination \cite{heiranian2015water}, and the formation of clathrate hydrates \cite{loveday2001transition,schaack2019observation}. The hydrogen bonds between water molecules induce complex structural motifs, resulting in more than 17 ice polymorphs in nature \cite{salzmann2019advances}. Atomic-level investigations of ice polymorphs have been intensively studied by molecular dynamics (MD) simulations, where various molecular fingerprints have been proposed for phase identification based on the positions and momenta of atoms in the simulation. Unlike simple elemental phases such as FCC, BCC, and HCP, the molecular symmetries in ice phases are much more complex. For example, 46 water molecules are included in a unit cell of sI ice, in contrast to four in an FCC unit cell. Furthermore, the inevitable thermal fluctuations at finite temperature make the classification of ice phases even more challenging. Considerable efforts have been devoted to designing accurate atomic fingerprints for the classification of water molecules, which can be divided into two categories: 1) traditional order parameters based on domain knowledge and 2) machine learned (ML) order parameters.
	
	Traditional parameters, such as Voronoi polyhedra \cite{ruocco1992analysis}, bond-orientational order parameter \cite{steinhardt1983bond, lechner2008accurate},  tetrahedral order parameter \cite{chau1998new,errington2001relationship},and local-structure index \cite{shiratani1996growth}, have been widely used in studies of ice nucleation, free energy calculations, and liquid-liquid transitions. For instance, the Voronoi polyhedra method recognizes the local ordering structure by generating polyhedra with topological properties, which are used to analyze the local structure of liquid water \cite{yan2007structure}. The bond order parameter investigates the molecular structure based on spherical harmonics by introducing different sets of order parameters designed for specific properties of water phases \cite{steinhardt1983bond, lechner2008accurate}. These types of order parameters are built with hand-crafted mathematical expressions for specific molecular arrangements of interest, which heavily rely on the expert's intuition and usually capture partial geometric information of structures. Therefore, for the classification of unexpected or complex systems with phase coexistence, an optimized combination of multiple sets of order parameters or newly designed ones will be desired, but this is time-consuming and may potentially introduce systematic biases to phase classification.
	
	Machine learned order parameters aims to overcome some limitations of traditional ones by leveraging the latent features learned by deep neural networks to distinguish different water phases. Early attempts include the dense network-based classification method proposed by Geiger and Dellago \cite{geiger2013neural}, which uses a set of symmetry functions to characterize the local atomic environment; the DeepIce network developed by Fulford \textit{et al}. \cite{fulford2019deepice}; and the PointNet architecture proposed by DeFever \textit{et al}. \cite{pointNN}, which directly operates on the atomic positions without system-specific feature engineering. More recently, graph neural networks have gained popularity due to their ability to naturally represent atomic systems of flexible sizes, with atoms as nodes and bonds as edges.  ML typically requires large training database, such as MD trajectories of different water phases, and careful tuning of weights and hyperparameters to minimize prediction errors, particularly for structures outside the training dataset \cite{Monserrat2020NC}. Additionally, supervised ML requires preparation of training labels. Despite the excellent classification accuracy achieved by these methods, notable misclassifications still exist. Misclassification errors often stem from the ambiguous distinction of latent feature spaces between different water phases. For instance, GCIceNet misclassified some atoms from ice-II, ice-III, and ice-VI phases due to the partially overlapping latent features of the three phases \cite{icenet}. Such overlapping is largely attributed to the thermal perturbations in the atomic system introduced in the molecular dynamics simulation. Thus, the removal of thermal noise has been regarded as an effective preprocessing step for MD simulation trajectories to distinguish the feature spaces of various phases and facilitate phase classification \cite{Hsu-denoiser}.
	
	Building upon our previous work based on generative artificial intelligence (genAI) to identify bulk and defect structures with simple symmetries \cite{Hsu-denoiser}, we develop a score-based denoiser model to remove the thermal noise of water molecular polymorphs (including seven fully ordered ice phases and one liquid water phase) formed by a complex hydrogen network. This is followed by a model-free classification method that identifies the phases by comparing the structural similarity of denoised structures with ideal reference phases. Our unsupervised phase classification framework has two key highlights: 1) The unsupervised training of the denoiser model  requires merely ideal ice phases, in contrast to data-intensive ML approaches that need the preparation of large trajectory datasets by running molecular dynamics simulations with carefully chosen force fields and/or pre-labeling efforts of water phases. 2) In the model-free classification step, instead of applying hand-crafted order parameters or machine-learned latent features, we employ a universal atomic descriptor, the Smooth Overlap of Atomic Positions (SOAP) \cite{SOAP}, which ensures Euclidean symmetries with translation and rotation invariance and transferability to any structural system with different symmetries. This approach allows us to distinguish ice phases in the test datasets with a remarkable 100\

	\section{Method}
	\label{sec:headings}
	
	\subsection{Denoiser model}
	As discussed in Ref.~\cite{Hsu-denoiser}, the key component of the denoiser model is the noise prediction network $\varepsilon_{\theta}(\mathbf{r})$ that predicts the ``noise'' or displacement vectors of input atomic structure $\mathbf{r}$ with respect to reference structure $\mathbf{r}_0$. Such a model can be used to ``denoise'' or remove noise from the input $\mathbf{r}$ to produce a less noisy structure $\mathcal{D}(\mathbf{r})=\mathbf{r}-\varepsilon_{\theta}(\mathbf{r})$, and iterative applications of the denoising operator $\mathcal{D}$ yield ``cleaner'' structures $\mathcal{D} \circ \dots \mathcal{D}(\mathbf{r})$ that are closer to $\mathbf{r}_0$, where ``$\circ$'' designates function composition. According to Vincent \cite{Vincent2011scorematching}, $\varepsilon_{\theta}$ can be trained using the denoising score matching method (Algorithm \ref{alg:Denoiser}), which has been extensively utilized in generative AI approaches such as the diffusion model \cite{Sohl-Dickstein2015-DPM, Ho2020-DDPM, Song2020-unified}. A generative or probabilistic explanation of this algorithm links denoising $\varepsilon_{\theta}(\mathbf{r}) = -\sigma s_{\theta} (\mathbf{r})$ to the score function defined as the partial derivative of data distribution $P(\mathbf{r})$ with respect to data (atomic coordinates here) $s(\mathbf{r})=\partial P(\mathbf{r})/\partial \mathbf{r}$. Here  $P(\mathbf{r})$ is the kernel density estimation of the training set of $m=7$ reference structures $\{\mathbf{r}_0^{(1)}, \dots, \mathbf{r}_0^{(m)}\}$ \cite{Vincent2011scorematching, Hsu-denoiser}, and $\sigma$ is the chosen noise magnitude of Gaussian kernels during training. The score is the gradient vectors on each atom, analogous to atomic forces, that point towards the clean reference atomic positions.
	Intuitively, the probabilistic score-based denoising process is not dissimilar to energy minimization, which can be regarded as denoising a structure by minimizing energy $U(\mathbf{r})$ or maximizing Boltzmann distribution $P\propto \exp(-U/k_B T)$. Both methods move atoms to make a structure more “plausible”, with scores playing the same role as forces in energy minimization. However, our approach learns a probability distribution from a simple reference dataset without requiring an energy model.

	The noise prediction model was approximated with the NequIP  \cite{nequip}
	
	model, which achieves rotational equivariance 
	
	through weighted tensor products between the irreducible representations of input features, i.e. atomic coordinates and auxiliary information (atomic number), and spherical harmonics representations of bond vectors \cite{nequip}. With the score model, perturbed input configurations tend to be denoised to resemble the references. 
	
	A notable advantage of the denoiser approach is that structures physically distinct from the training set are not unduly forced into the references:
	case studies on  FCC, BCC, HCP structures demonstrate that physical 3D/2D/1D/0D defects such as melt, grain boundaries,  dislocation lines, interstitials and vacancies are clearly revealed, while crystalline phases are classified with 100\

	Following our previous work \cite{Hsu-denoiser}, the denoiser model is trained using entirely synthetic data generated from reference data consisting of ideal structures of seven ice bulk supercells (Ice-Ih, Ice-Ic, Ice-II, Ice-III, Ice-VI, Ice-VII, sI-hydrate) constructed using the GenIce software \cite{Matsumoto2018JCC}. Our current method is limited to ordered structure without partial occupancy or disorder. During training, super-cells replicated from the unperturbed ice unit-cells were sampled  with a batch size of 8, with i.i.d Gaussian noises of amplitude $\sigma$  added to atomic coordinates. We employ the AdamW optimizer \cite{AdamW} with a learning rate of $2 \times 10^{-4}$ over 300,000 training updates. Detailed information on model training and the selection of hyperparameters is provided in Algorithm \ref{alg:Denoiser}.
	
	The maximum noise level $\sigma_{\text{max}}$  was chosen to be 13\

	\begin{algorithm}
		\caption{Unsupervised denoiser training}
		\label{alg:Denoiser}
		\SetAlgoNlRelativeSize{0}
		\SetAlgoNlRelativeSize{-1}
		\SetAlgoNlRelativeSize{1}
		\SetAlgoNlRelativeSize{-1}
		\SetAlgoNlRelativeSize{0}
		\SetKwInput{KwInput}{Input}
		\SetKwInput{KwOutput}{Output}
		
		\KwInput{ 
			Training dataset $\{(\mathbf{r}, \mathbf{z})\}$ (coordinates and atomic numbers), number of training steps $T$,

			score model $\varepsilon_{\theta}$, 
			
			optimizer $\text{Opt}$, and learning rate $\eta$}
		\KwOutput{$\varepsilon_{\theta}$ with optimized $\theta$}
		\For{$t \leftarrow 1$ \KwTo $T$}{
			
				$(\mathbf{r}, \mathbf{z}) \sim \{(\mathbf{r}, \mathbf{z})\}$  \\

				$\sigma \sim \mathcal{U}(0,\sigma_{\text{max}})$, $\epsilon \sim \mathcal{N}(0,I)$\\
				$\delta \leftarrow  \sigma\epsilon$ \\         
				$\hat{\delta} \leftarrow  \varepsilon_{\theta}({r}+\delta, z)$\\
				
				$L \leftarrow   \|\delta - \hat{\delta}\|^2  $\\
				${\theta} \leftarrow \text{Opt}(L, \theta, \eta)$
				
		}
	\end{algorithm}

	\subsection{Unsupervised training of classifier}
	
	The test dataset for the  classifier consists of molecular dynamics (MD) trajectories of seven ice phases: Ice-Ih, Ice-Ic, Ice-II, Ice-III, Ice-VI, Ice-VII, sI-hydrate, the liquid water phase, and the ice-water interface of Ice-Ih. The trajectories of the seven ice polymorphs and liquid water phase were obtained from Ref. \cite{icenet}, which contains 96734 water molecules prepared using MD simulation with the TIP4P/Ice water model. The preparation of MD trajectories for the ice-water interface is described in the Molecular Dynamics Simulation section. 
	
	We focus on the classification of different ice polymorphs using positions of heavy oxygen atoms only, which were found sufficient. To characterize the local environment of each O atom, SOAP descriptors \cite{SOAP} are computed utilizing the DScribe package \cite{dscribe}. To ensure accurate representation of atomic environments across all ice phases, we utilized 12 radial basis functions and a maximum degree of 10 for spherical harmonics in the calculation of SOAP descriptors. Standard scalar transformation and principal component analysis (PCA) transformation, retaining 30 principal components, were employed for standardization and dimensionality reduction. Then, we calculated the similarity distance (Euclidean or cosine distance) between the reduced SOAP descriptors of each atom in the test trajectories and those in the seven ideal phases, as described in Algorithm \ref{alg:classifer}. The minimum distance indicates the most likely phase type for the test atom. Details on the parameter selections for SOAP descriptors and PCA components can be found in SI.

	The liquid phase encompasses a complicated spectrum of structural motifs \cite{Monserrat2020NC} as its complex hydrogen network may dynamically alter its local arrangements. To simplify the classification task, we assume that any atom not uniquely identified as one of the seven ice phases will be classified as the liquid water phase. The threshold for differentiating between the solid and liquid phases can be determined from the distribution of the similarity score $ \min_{y \in Y} S(x, y)$ between the features $x$ of the atoms in denoised ice and liquid water phases and the seven ideal reference structures $Y$. The results will be discussed in the next section.

	Any atom with a similarity score larger than the threshold value will be regarded as the liquid water phase in the classification algorithm. Otherwise, they will be assigned to the ice phase type with the minimum similarity distance.
	
	\begin{algorithm}
		\SetKwInput{KwInput}{Input}
		\SetKwInput{KwOutput}{Output}
		\SetAlgoLined
		\caption{Classification}
		\label{alg:classifer}
		
		\KwInput{
			Structure $\mathbf{r}$,
			unit-cells of seven ideal ice phases $\{\mathbf{r}_0^{(i)}\}$, descriptor function $F$, similarity measure $S$ (Euclidean or cosine),
			threshold of similarity distance $\delta$
		}
		\KwOutput{Label $\text{phase}(x)$ for each atom $x$ in $\mathbf{r}$}
		
		\For{ $i \in [1 \dots 7]$, $y \in \{\mathbf{r}_0^{(i)}\}$}{
			$ f_{i,y} =   F(\mathbf{r}_0^{(i)}, y) $\\
		}
		\For{$ x \in \mathbf{r}$}{
			$ f_x = F(\mathbf{r}, x) $ \\

			\eIf{\( \min_{i,y } S(f_x, f_{i,y}) > \delta \)}{
				$ \text{phase}(x) = \text{liquid water} $ 
			}{
				\( \text{phase}(x) = \text{argmin}_{i } S(f_x, f_{i,y}) \)
			}
		}
	\end{algorithm}

	\subsection{Molecular dynamics simulation}
	
	The MD trajectories of ice-water interfaces were prepared using NPH (constant number of particles, pressure, and enthalpy) simulations with the LAMMPS molecular dynamics package \cite{LAMMPS}. The system sizes for the Ice-Ih/liquid, Ice-Ic/liquid, and Ice-sI/liquid interfaces were 24,000, 24,000, and 37,260 atoms, respectively. The simulation consisted of several stages: (1) NVT (constant number of particles, volume, and temperature) equilibration at 250 K for 10,000 steps; (2) NPT (constant number of particles, pressure, and temperature) equilibration at 250 K for 20,000 steps; (3) NPT heating to slightly below the melting temperature (267 K according to the interatomic potential used \cite{cheng2019ab}) for 20,000 steps; (4) NPT melting of half of the box for 20,000 steps; (5) NPT cooling of the liquid region to just above the melting temperature for 20,000 steps; (6) NPT equilibration to resolve any anisotropic stress for 20,000 steps; and (7) NPH production run for 100,000 steps. A timestep of 0.25 fs was used throughout the simulations, and the pressure was maintained at 1 bar. The BPNN water interatomic potential \cite{cheng2019ab} was employed to accurately model the water molecules in the system.

	\section{Results and discussion}
	Before delving into the detailed results, we note that our denoiser model and the classifier are solely based on the ideal reference structures with no exposure to test datasets, including those from previous publications and our own MD simulations.
	
	\subsection{Denoising}
	\begin{figure}[ht]
		\begin{center}
			\includegraphics[width= 0.95\columnwidth]{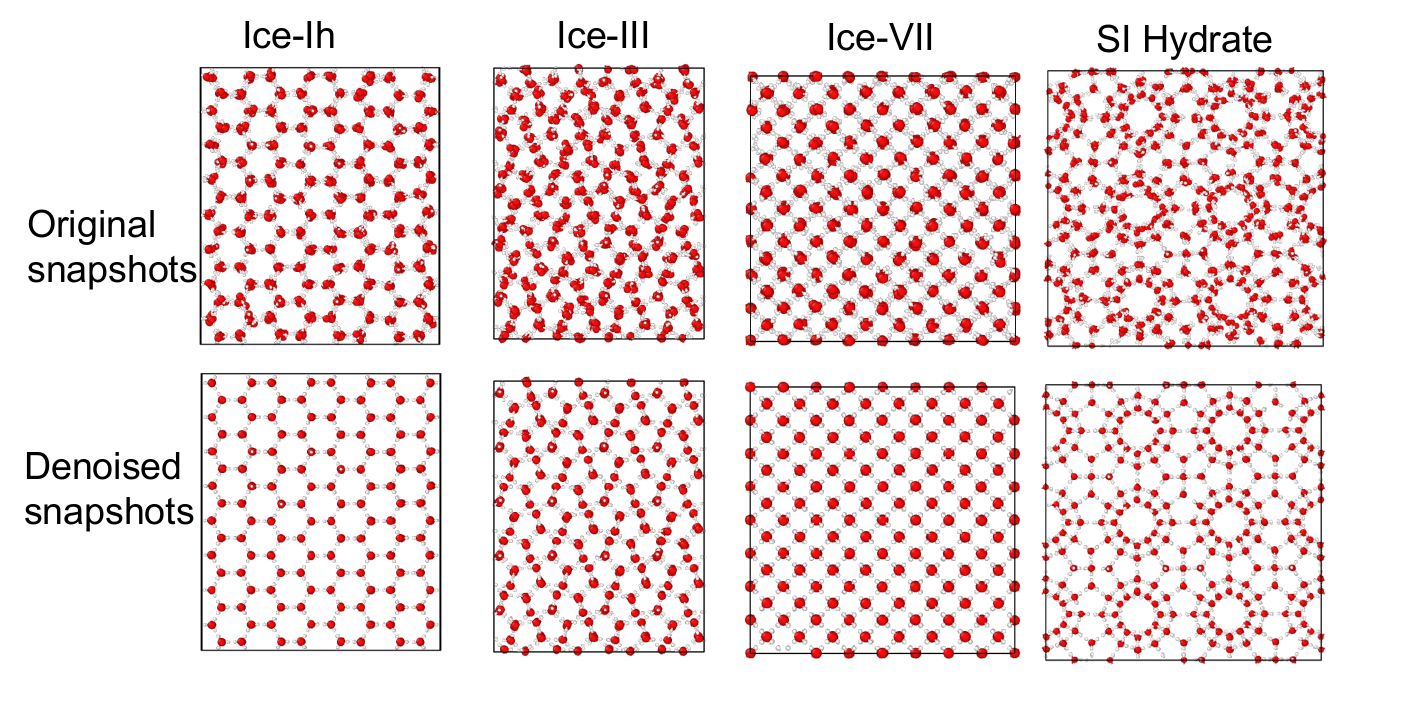}
		\end{center}
		\caption{MD snapshots depicting four ice bulk phases (ice-Ih, ice-III, ice-VII, and SI Hydrate) before and after denoising. The additional three bulk phases (ice-Ic, ice-II, and ice-VI) are displayed in Figure S1 (supporting information).}
		\label{fig:snapshots}
	\end{figure}
	Our prior research \cite{Hsu-denoiser} has demonstrated the capability of score-based denoising models in accurately identifying phase transitions for material systems with simple lattice symmetries, such as face-centered cubic (FCC), hexagonal close-packed (HCP), body-centered cubic (BCC), and $\beta$-quartz. However, the denoising capability of the score-based model has not been evaluated for molecular systems with complex symmetries. In the case of water phases, hydrogen bonding gives rise to diverse packing arrangements of water molecules, resulting in more than 17 ice polymorphs \cite{salzmann2019advances}. For example, the unit cell of the SI hydrate ice phase contains 46 water molecules, which is significantly more complex compared to the 2 and 4 unique atom sites in BCC and FCC unit cells, respectively. This complexity poses a challenge for achieving a perfect resemblance to the ideal Wyckoff positions for a thermally perturbed ice structure within the gradient field of a score model. In this study, we demonstrate that by training the score-based denoiser model described in Algorithm \ref{alg:Denoiser} with randomly perturbed referenced phases, we can effectively remove thermal noise of MD trajectories of ice phases within 5-8 iterative denoising steps. This process enables us to recover the pristine bulk phases from MD snapshots  at finite temperature. Figure \ref{fig:snapshots} presents a comparison of MD snapshots for four different ice phases: ice-Ih, ice-III, ice-VII, and SI Hydrate, before and after denoising. Figure S1 showcases the three additional ice bulk phases (ice-Ic, ice-II, and ice-VI). It is noteworthy that denoising the liquid water phase results in nearly all atoms remaining disordered, highlighting the model's ability to distinguish between ordered and disordered phases.
	
	The effectiveness of removing thermal noise to facilitate ice phase identification can also be observed by examining changes in their feature representations. Unlike the straightforward application of orientation order parameters like Steinhardt features $\bar{q}_4$ and $\bar{q}_6$, or Common Neighbor Analysis (CNA) for metal systems, ice phases with complex symmetries demand more sophisticated descriptors to capture their atomic environments accurately. We utilized Smooth Overlap of Atomic Positions (SOAP) descriptors \cite{SOAP}, which encode atomic geometries using a local expansion of Gaussian smeared atomic density with orthonormal functions based on spherical harmonics and radial basis functions. This method allows for the rapid extraction of detailed atomic geometry information directly from the material system's atomic coordinates with prior application to quantify liquid water and ice structures \cite{Monserrat2020NC}.

	\begin{figure}[ht]
		\begin{center}
			\includegraphics[width= 0.9\columnwidth]{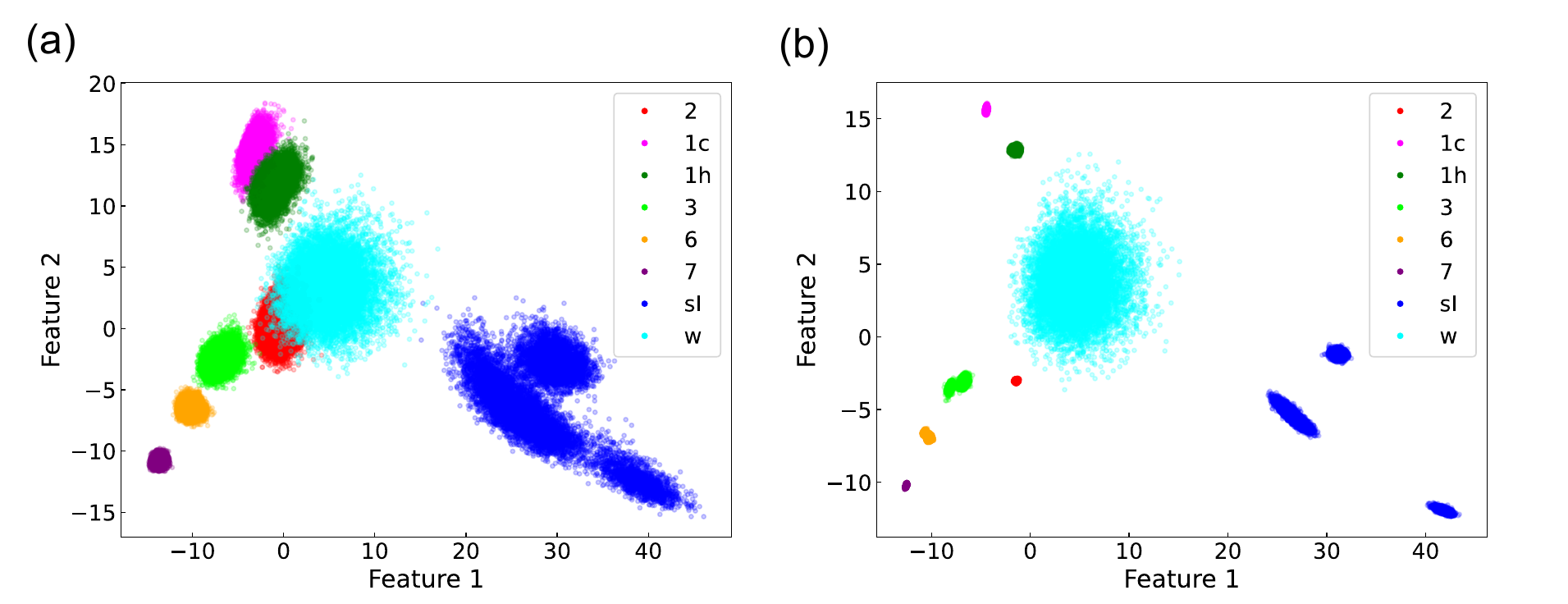}
		\end{center}
		\caption{ Distributions of two-dimensional latent variables obtained with PCA of SOAP descriptors for seven ice bulk phases before (a) and after denoising (b). Dots are colored according to the structure types: ice-II (red), Ic (magenta), Ih (dark green), III (bright green), VI (orange), VII (purple), SI (blue) and water (cyan)}.
		\label{fig:feature_map}
	\end{figure}
	
	Figure \ref{fig:feature_map} illustrates the distribution of two principal components of SOAP features in MD trajectories of seven ice bulk phases and liquid water before (a) and after (b) denoising. Upon applying the denoiser model, the distribution of per-atom features in each solid phase  becomes sharper and more concentrated. Fig.~\ref{fig:feature_map}b reveals well separated distributions, in contrast to substantial overlap observed among ice-Ic and ice-Ih, ice-III, ice-II, and liquid water in Fig.~\ref{fig:feature_map}a before denoising. The significant reduction of thermal noise results in a clearer SOAP descriptor space across all MD trajectories, thereby facilitating subsequent classification tasks. Note that one structure may contain more than one Wyckoff positions and corresponding sets of 2D latent variables, e.g. blue points for sI in Fig.~\ref{fig:feature_map}.  Note also that the feature distribution may not converge into discrete dots entirely due to residual thermal noise not completely removed. While the trained denoiser ML model is effective, it may not achieve perfect ideal structure recovery.

	Figure \ref{fig:distribution-similarity} depicts the statistical distribution of the calculated similarity scores $ \min_{y \in Y} S(x, y)$ (details in Algorithm \ref{alg:classifer})  between the features $x$ of the atoms in denoised trajectories of ice and liquid water phases and the seven ideal reference structures $Y$. The histogram plot of the similarity scores for all atoms reveals two distinct distribution patterns. According to their phase labels, the leftmost distribution curve (orange {in \ref{fig:distribution-similarity}a), which is sharply concentrated around 0, represents all ice phases, while the atoms falling within the wide Gaussian distribution are entirely from the liquid water phase (blue). The middle position of the valley boundary that divides the two distribution patterns can be chosen as the threshold for differentiating between the solid and liquid phases. The cosine similarity measure (Fig.~\ref{fig:distribution-similarity}a) was chosen over Euclidean distance (Fig.~\ref{fig:distribution-similarity}b) for the classifier.
		
		Any atom with a similarity score larger than the threshold value will be regarded as the liquid water phase in the classification algorithm. Otherwise, they will be assigned to the ice phase type with the minimum similarity distance.

		The findings of this work reaffirms the hypothesis that a denoising model training on synthetic Gaussian noises can successful predict and remove physical thermal fluctuations \cite{Hsu-denoiser}.
		\begin{figure}[ht]
			\begin{center}
				\includegraphics[width= 0.95\columnwidth]{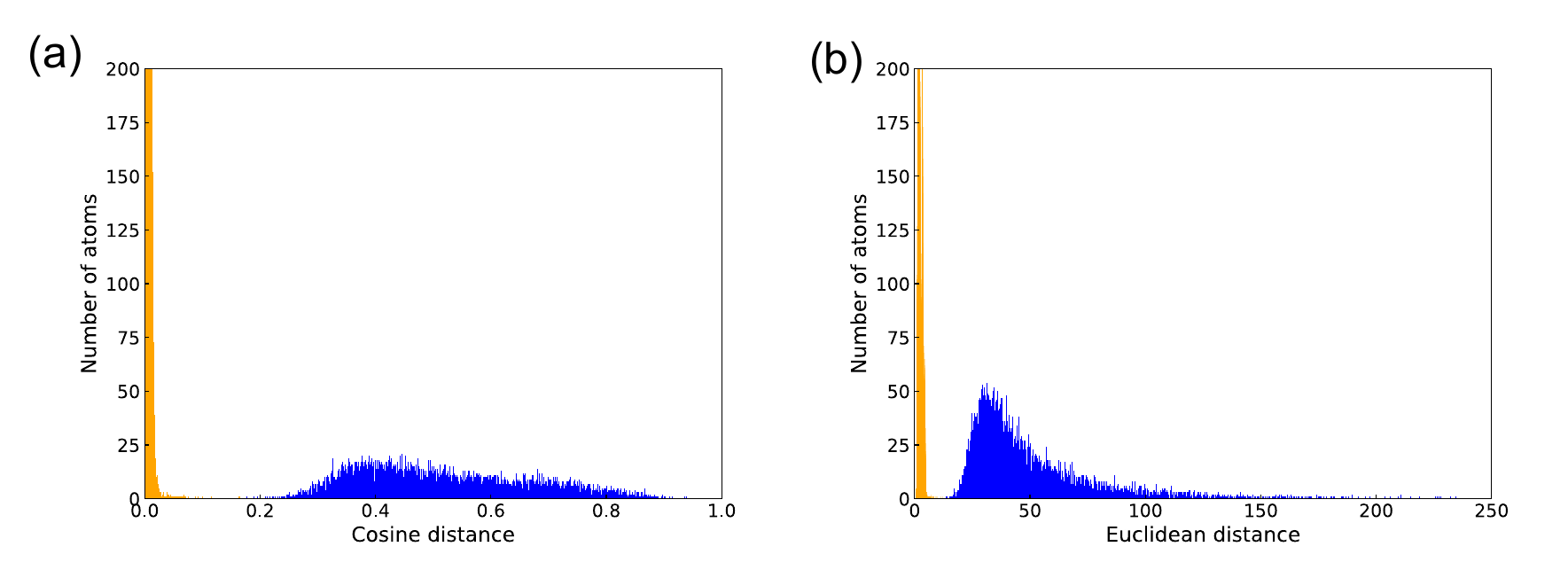}
			\end{center}
			\caption{Distribution of similarity scores $ \min_{y \in Y} S(x, y)$ of input atom features $x$ against features $y$ from ideal reference structures $Y$, measured using cosine distance (a) and Euclidean distance (b). Atoms in denoised bulk ice polymorphs and in liquid water are shown as orange and blue, respectively.}
			\label{fig:distribution-similarity}
		\end{figure}
		
		\subsection{Classification}
		
		The classification of all ice phases is achieved for MD trajectories using a model-free classification method outlined in Algorithm \ref{alg:classifer}, obviating the need for further training efforts. We evaluated our classification accuracy across all MD trajectories of seven ice bulk phases and compared the results with other water phase classification methods listed in Table \ref{table:accuracies}. Remarkably, our classification method achieved a perfect accuracy of 100\
		
		It is important to note that there are differences in the test datasets of our model compared to the models reported in other references listed in Table \ref{table:accuracies}. Our classifier model of 100\
		On a fundamental level, classification is a subjective matter with no perfect solution that fits all. When it comes to crystalline phases with very small thermal perturbations, it can be considered unambiguous, which is exactly the route we took to denoise an input structure as a preprocessing step before attempting classification. However, to our knowledge there is no precise and unambiguous method of classifying liquid phases.
		
		Thus, to ensure a reliable classification, we only performed denoising over the test atoms of seven ice polymorphs with well-studied symmetries. Despite this limitation, our unsupervised classification with 100\
		
		\begin{table*}[t]
			\begin{center}
				\begin{tabular}{ ccccccccc }
					\hline
					\hline
					Unsupervised & PCA \cite{icenet} & Autoencoder \cite{icenet} & CGIceNet \cite{icenet} & SOAP & Denoiser \\
					
					&74.2& 85.8 & 98.3 & 90.2 & 100\\
					\hline
					\hline
					Supervised & SVM \cite{icenet} & RF \cite{2023topology_gnn} &Top2Phase \cite{2023topology_gnn} & CGIceNet \cite{icenet} & PointNet \cite{pointNN} & DeepIce \cite{fulford2019deepice}\\
					& 92.7 & 89.2 &99.9 & 99.8 & 99.6& 99.6 \\
					\hline
					\hline
				\end{tabular}
				\caption{ 
					Classification accuracy of the bulk water phases with  unsupervised learning algorithms 
					and supervised learning algorithms}
				\label{table:accuracies}
			\end{center}
		\end{table*}
		\newpage
		
		Accurate and automatic identification of ice and liquid water phases in ice-liquid two-phase systems is crucial for studying the kinetics of melting transitions, ice growth from super-cooled water, and the ice/quasi-liquid interface when exposed to air. Here, we demonstrate the application of our classification method to MD trajectories of ice-liquid two-phase simulations during the melting of three distinct ice polymorphs, i.e., ice-Ic, ice-Ih, and ice-sI. Detailed MD simulation parameters related to the melting simulations can be found in the Methods section.
		
		\begin{figure}[h]
			\begin{center}
				\includegraphics[width= 0.9\columnwidth]{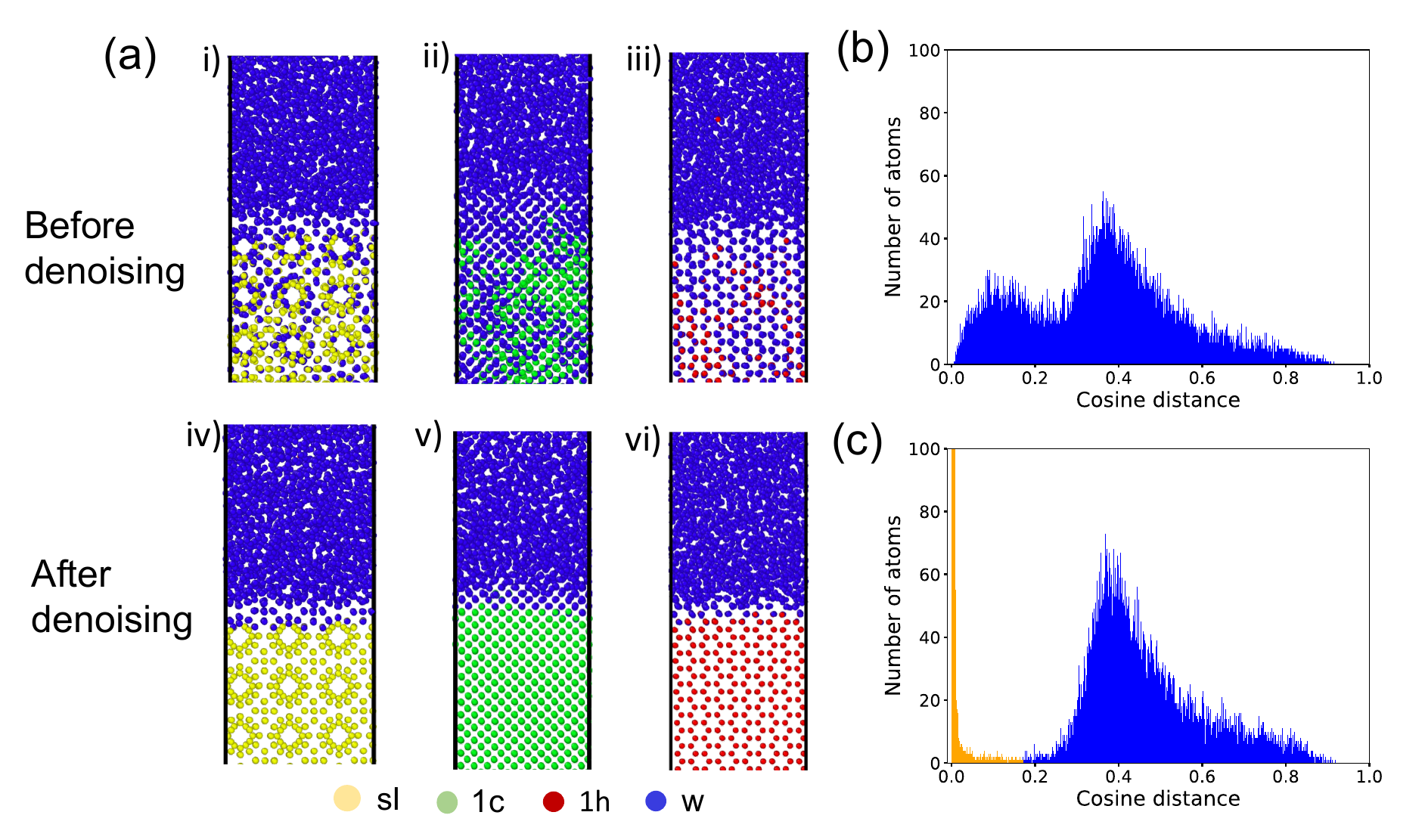}
			\end{center}
			\caption{ (a) MD snapshots of ice-liquid water interfaces before (i-iii) and after denoising (iv-vi). Atoms are color coded according to the identified phase by the classifier model. The distribution plots of similarity score of the classifier model for atoms in the three interface structures before (b) and after (c) denoising.}
			\label{fig:twophase}
		\end{figure}
		
		Figure \ref{fig:twophase} (a) presents snapshots of ice-water interface structures before (i-iii) and after (iv-vi) applying the denoiser model. The color coding represents the ice phase identified by calculating the similarity score using the classifier model (Algorithm \ref{alg:classifer}). Before denoising, the color coding shows a mixture of water and ice phase labels, indicating the presence of thermal fluctuations and disorder at the interface and the ice bulk regions. However, after denoising, the color coding shifts to solely ice phase labels in the bulk region, revealing the underlying crystalline structure. Distinct boundaries are observed at the liquid-ice interface, and the identified phase results perfectly agree with the phase labels in MD simulation, despite our unsupervised classification having no prior knowledge of any phase labels. The statistical distribution of cosine similarity scores of atoms in the three interface structures before and after denoising is depicted in Figure \ref{fig:twophase} (b) and (c), respectively. Before denoising, the similarity scores show two connected wide distribution patterns with peaks centered around 0.1 and 0.4, suggesting the presence of both liquid-like and ice-like local environments. After denoising, a sharp peak emerges around 0.0, indicating the emergence of ideal ice phases, while the liquid phase exhibits a broad Gaussian distribution spanning 0.2-0.9, similar to the distribution plot of ice phases and liquid phases shown in Figure \ref{fig:distribution-similarity}. The denoised snapshots and the successful classification of ice and liquid phases at the interface demonstrate the ability of our approach to remove thermal noise and cleanly reveal the intrinsic structural features of intricate bi-phase systems, which captures the underlying crystalline structure of the ice phase while preserving the liquid-like nature of the water phase. It underscores the potential of our denoising and classification method for investigating the complex behavior of ice-liquid interfaces and other systems exhibiting phase transitions at the atomic scale.

		\section{Conclusions}
		In conclusion, we have presented a novel phase classification framework that combines a score-based denoiser model with a model-free classification approach to accurately identify ice phases in molecular dynamics simulations. Our approach addresses the limitations of traditional order parameters, which often rely on expert intuition and encode partial geometric information, and machine learning-based methods, which require large training datasets. By leveraging the power of generative AI and score-based denoising model, thermal noise can be successfully removed from ice polymorphs in molecular dynamics simulations, facilitating the subsequent classification tasks that identify phases by comparing the structural similarity of denoised structures with ideal reference phases. The proposed approach achieves 100\
		
		Moreover, the successful application of our denoising and classification approach to ice-liquid two-phase systems highlights its versatility in revealing the underlying crystalline structure of the ordered phases while preserving the liquid-like nature of the disordered phase. It showcases the possibility to enable a detailed analysis of thermodynamics and kinetics of the melting process, ice growth from super-cooled water, and the behavior of the ice/quasi-liquid interface at the atomic level without the need for prior phase labeling. This makes it a powerful tool for investigating complex systems where phase boundaries and transitions may not be well-defined or easily identifiable. In summary, our phase classification framework represents an important advancement in the accurate identification of ice phases in molecular dynamics simulations. The simplicity, generalizability, and high accuracy of our approach make it a valuable tool for investigating structural evolution and phase identification in a wide range of materials. We believe that this work will stimulate further research and applications of score-based denoiser models in the field of phase classification, ultimately contributing to a deeper understanding of the complex behavior of water and other materials at the atomic scale.
		
		\section*{Author Contributions}
		F.Z. and V.L. secured funding. F.Z. conceived the idea, supervised the study. F.Z and T.H. constructed the code framework. H.S. performed the model training, code development, and analysis. S.H. prepared the testing dataset by molecular dynamics simulations. All authors contributed to discussions, analysis, and writing the paper.
		
		\section*{Acknowledgment}
		
		This work was performed under the auspices of the U.S. Department of Energy by Lawrence Livermore National Laboratory under Contract DE-AC52-07NA27344. This work was funded by the Laboratory Directed Research and Development (LDRD) Program at LLNL under project tracking code 22-ERD-016 and 23-SI-006. Computing support for this work came from the Lawrence Livermore National Laboratory institutional computing facility. IM release number: LLNL-JRNL-862778-DRAFT.
		
		\section*{Data and Software Availability }
		The source codes for training and deploying the denoiser model and the classifier model are freely available at \href{https://github.com/LLNL/graphite/tree/main/notebooks}{https://github.com/LLNL/graphite/tree/main/notebooks}. MD simulation data will be available on reasonable request.
		
		\section*{Supporting Information Available}
		Additional information about the effects of denoising and computational parameter choices can be found in Supporting Information.
		
		\newpage

		\section*{References}

		\bibliography{references}

	\end{document}